\documentclass[twocolumn,showpacs,preprintnumbers,amsmath,amssymb,superscriptaddress]{revtex4}

\usepackage{graphicx}
\usepackage{dcolumn}
\usepackage{bm}

\begin{document}


\title{Resonances of the Quantum $\delta$-Kicked Accelerator}

\author{V.\thinspace Ramareddy}
\affiliation{%
Department of Physics, Oklahoma State University\\
Stillwater, OK 74078-3072}
\author{G.\thinspace Behinaein}%
\affiliation{%
Department of Physics, Oklahoma State University\\
Stillwater, OK 74078-3072}
\author{I.\thinspace Talukdar}%
\affiliation{%
Department of Physics, Oklahoma State University\\
Stillwater, OK 74078-3072}
\author{P.\thinspace Ahmadi}%
\affiliation{%
Department of Physics, Oklahoma State University\\
Stillwater, OK 74078-3072}
\affiliation{School of Physics, Georgia Institute of Technology, Atlanta, GA 30332-0430}
\author{G.\thinspace S.\thinspace Summy}
\affiliation{%
Department of Physics, Oklahoma State University\\
Stillwater, OK 74078-3072}

\date{\today}

\begin{abstract}
We report the observation of  high order resonances of the quantum
$\delta$-kicked accelerator using a BEC kicked by a standing wave of
light. The signature of these resonances is the existence of quantum
accelerator modes. For the first time quantum accelerator modes were
seen near 1/4 and 1/3 of the Talbot time. Using a BEC enabled us to
study the detailed structure of the modes and resonances which are
related to the fractional Talbot effect. We present a general theory
for this system and apply it to predict the behavior of the
accelerator modes.

\end{abstract}

\pacs{05.45.Mt, 03.75.Lm, 32.80.Lg, 42.50.Vk}
\maketitle
Studies of the dynamics of cold atoms subject to momentum kicks from
a standing wave of off-resonant light have played an important role
in the development of atom optics and quantum chaos. This system has
been used in realizing the quantum $\delta$-kicked rotor
\cite{delta_kick} and the quantum $\delta$-kicked accelerator
\cite{newone,gil_prl1}. These experiments have allowed the
observation of important time asymptotic behavior such as dynamical
localization and quantum resonance \cite{dynamical}. Other
interesting classes of kicked systems studied theoretically include
the kicked harmonic oscillator \cite{HarOsc} and kicked Harper model
\cite{harper}. In the case of the $\delta$-kicked rotor, when the
kicking period is close to integer multiples of the primary
resonance time (the half-Talbot time), quadratic growth of energy is
observed. When a linear potential such as the one produced by the
Earth's gravity is added to the kicked rotor Hamiltonian, a kicked
accelerator is produced. In this case, when the pulse period is
close to a resonance time, symmetry is broken and a fixed momentum
per kick can be imparted to the atoms in a specific direction. Atoms
which are kicked in such a way are said to be in a Quantum
Accelerator Mode (QAM) \cite{gil_prl1}. Fishman, Guarneri and
Rebuzzini  (henceforth referred to as FGR) developed a framework
called $\epsilon$-classical theory \cite{qam_th}, where a parameter
$\epsilon$ proportional to the difference of the pulse period from
one of the primary resonances
plays the role of Planck's constant. In the limit of pulse period
very close to a primary resonance time, ($\epsilon\rightarrow 0$), a
classical mapping can be used to study the system. In addition to
the main QAM discovered in the first experiment \cite{gil_prl1}, the
FGR theory predicted the existence of  additional QAMs which were
subsequently observed by Schlunk {\it et al.} \cite{gil ho}. More
recently, it was found that the QAM can be understood in terms of
mode locking and an Arnol'd tongue analysis \cite{arnold tongue}.

Higher order quantum resonances of the $\delta$-kicked rotor have
been predicted to occur whenever the kicking period is a rational
fraction of the primary resonance time \cite{dokl}.  Recently such
higher order quantum resonances were observed using a kicked BEC of
Na atoms \cite{bil1} and using a thermal sample of atoms
\cite{canada}. Therefore it is interesting to ask the question: do
such resonances also exist in the quantum $\delta$-kicked
accelerator and can QAMs be formed in their vicinity? In this Letter
we report the observation of these higher order resonances and their
associated
 QAMs. We also show
how $\epsilon$-classical theory can be generalized to predict the
behavior of QAMs at these resonances. In addition, by treating the
standing light wave as a diffraction (phase) grating and using a
picture analogous to the fractional Talbot effect in optics
\cite{optics talbot}, we are able to explain the internal momentum
state (diffraction order) structure of the QAM.

To begin discussion of QAMs around higher order resonances, we write
the Hamiltonian of the quantum $\delta$-kicked accelerator in
dimensionless units as \cite{qam_th},
\begin{equation}
H=\frac{p^2}{2}-\frac{\eta}{\tau}x
+\phi_d\cos(x)\sum_{t}\delta(t^\prime-t\tau),\label{hamiltonian}
\end{equation}
where $p=n+\beta$ is momentum in units of two photon recoils ($\hbar
G$), $G=4\pi /\lambda$, $\lambda$ is the wavelength of the kicking
light, $n$ is the integer part of $p$, $x$ is the position in units
of $G^{-1}$, $\eta=Mg^\prime T/(\hbar G)$, $g^\prime$ is the
acceleration experienced by the atoms during the time between kicks
in the direction of the standing wave, $T$ is the kick period,
$\phi_d$ is the phase modulation depth (and represents the kicking
strength), and $M$ is the atom's mass. The scaled kick period is
$\tau=2\pi T/T_{1/2}$ where $T_{1/2}=2\pi M/(\hbar G^2)$ is referred
to as the half Talbot time and $t^\prime$ is the continuous time
variable in units of $T_{1/2}/2\pi$. In previous work
\cite{gil_prl1,newone} it was found that the QAM occur whenever $T$
is close to an integer multiple of $T_{1/2}$. It should also be
noted that in the frame that is accelerating with $g^\prime$, the
quasi momentum $\beta$ (fractional part of the momentum) is
conserved. For primary resonances FGR define a parameter
$\epsilon=\tau-2\pi l=2\pi(T/T_{1/2}-l)$, where $l$ is an integer.
The $\epsilon$ represents the closeness of the kick period to one of
the primary resonance times. This enables the one kick evolution
operator to be written as $U_\beta(t)={\rm e}^{-{i \over
|\epsilon|}{\tilde k} \cos\hat \theta}{\rm e}^{-{i \over
|\epsilon|}\hat H_\beta},$ where $\hat H_\beta = \frac{1}{2}
\frac{\epsilon}{|\epsilon|}\hat I^2 + \hat I\left[\pi l^\prime+\tau
\left(\beta+t\eta+\eta/2\right)\right]$, $\hat I=-i|\epsilon|{{\rm
d}\over {\rm d}\theta}$, $\tilde{k}=|\epsilon|\phi_d$, $\theta=x\
{\rm mod}\ 2 \pi$, and $l^\prime$ is an integer. By analogy with the
kicked rotor, higher order resonances are expected when $T$ is a
rational fraction $(a/b)$ of $T_{1/2}$ \cite{dokl,bil1}, where $a$
and $b$ are integers. Thus we now define $\epsilon=\tau-2\pi a/b$,
which represents the closeness parameter to one of the higher order
resonance times. Using this $\epsilon$ and the condition that at
higher order resonances $\left(m^2+2ml^\prime/a \right)a/2b$ is an
integer, where $m$ is an integer \cite{bil1,dokl}, the evolution
operator becomes
\begin{equation}
U_\beta(t)={\rm e}^{-{i \over |\epsilon|}{\tilde k} \cos\hat
\theta}{\rm e}^{-{i \over |\epsilon|}\hat H_\beta^h},
\label{operatorh}
\end{equation}
where $\hat H_\beta^h=\frac{1}{2}\frac{\epsilon}{|\epsilon|}\hat I^2
+ \hat I \left(\pi l^\prime/b+\tau\left(\beta+t\eta
+\eta/2\right)\right)$. Since $|\epsilon|$ plays the role of
Planck's constant, in the limit $\epsilon \rightarrow 0$,
Eq.~(\ref{operatorh}) produces the following classical mapping:
\begin{eqnarray} J_{t+1}
&=& J_{t}+{\tilde
k}\sin(\theta_{t+1})+\frac{\epsilon}{|\epsilon|}\eta\tau \nonumber \\
\theta_{t+1} &=& \theta_t+\frac{\epsilon}{|\epsilon|}J_t,
\label{map}
\end{eqnarray}
where $J_t$ is defined as
$J_t=I_t+\frac{\epsilon}{|\epsilon|}\left(\pi
l^\prime/b+\tau\left(\beta+t\eta +\eta/2\right)\right)$. The mapping
of Eqs.~(\ref{map}) is very similar to that found for the primary
resonances \cite{qam_th}. Hence the $\epsilon$-classical theory can
be applied to QAM appearing at higher order resonance times. This
gives the average momentum transferred to a QAM (in units of $\hbar
G$) near $(a/b)\ T_{1/2}$ as
\begin{equation}
\bar{p}_{\rm QAM}=-t{\eta \tau \over \epsilon}.  \label{momentum}
\end{equation}
Thus the general signature of a quantum resonance of the quantum
$\delta$-kicked accelerator is expected to be the asymptotic
divergence of a QAM's momentum, $\bar{p}_{\rm QAM}$, to infinity as
$\epsilon\rightarrow 0$, that is, when the kicking period approaches
the resonance time.

To experimentally observe these quantum resonances we subjected a
BEC to pulses of standing wave light as described in detail in
\cite{newone}. Briefly, the BEC was created in an optical trap and
consisted of approximately 30000 Rb-87 atoms in the $F=1$,
$5S_{1/2}$ level. After release from the trap, the BEC was kicked by
780 nm light which was 6.8 GHz detuned to the red of the atomic
transition. These parameters gave a value for the half-Talbot time
of $T_{1/2}=33.15\ {\rm \mu s}$. This light propagated through two
acousto-optic modulators (AOMs) to control the initial momentum of
the atoms with respect to the standing wave. This was accomplished
by driving the two AOMs with different frequencies. The kicking beam
was oriented at $48^\circ$ to the vertical making $g^\prime=6.6$
ms$^{-2}$. In order to vary the kicking strength $\phi_d$, the
length of the kicking pulses was adjusted. Typically the pulse
length was approximately 2.5 $\mu$s giving $\phi_d\approx1.5$. The
value of $\phi_d$ was estimated by comparing the relative population
of various diffraction orders after one kick. Note that the
population in the $l$-th order is given in terms of Bessel functions
via $|J_l(\phi_d)|^2$ \cite{qam_th_gil}. The momentum distribution
of the BEC was measured by taking an absorption image $\approx$ 8 ms
after the completion of the kicking sequence. Finally it should be
noted that the mean field energy was weak enough that it could be
ignored, making the Hamiltonian of Eq.~(\ref{hamiltonian}) a valid
approximation.

\begin{figure}[h]
  \includegraphics[width=8cm]{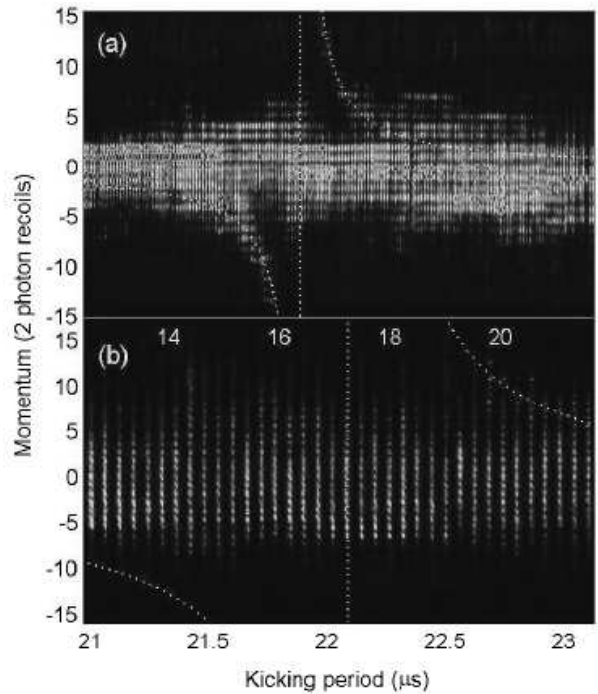}\\
  \caption{Horizontally stacked momentum distributions for different kicking periods for
  (a) 40 kicks across (1/2)$T_{1/2}$ ($b=2$) (b) 30 kicks across (2/3)$T_{1/2}$ ($b=3$).
  The initial momentum is chosen such that the part of the mode below the resonance time
  is populated more strongly in the case of (a) and vice versa in the case of (b).
  The dotted curve is predicted by the $\epsilon$-classical theory of Eq.~(\ref{momentum}).}\label{1by4period}
\end{figure}

Figure~\ref{1by4period} shows experimental scans of the kicking
period across two different higher order resonances. These figures
were generated by horizontally stacking the absorption images each
with a different kick period. The dotted curves are the QAM momenta
predicted by the $\epsilon$-classical theory of
Eq.~(\ref{momentum}). A  value for $\phi_d$ of 1.4 was used in
Fig.~\ref{1by4period}(a) near (1/2) $T_{1/2}$ and 1.8 in
Fig.~\ref{1by4period}(b) near (2/3) $T_{1/2}$. It can be seen that
the theory provides a good description of the momentum of the QAM
and confirms the presence of the higher order resonances. Both
experiments and numerical simulations \cite{sim} suggested that
higher values of $\phi_d$ were required to produce observable QAMs
as $b$ was increased.
\begin{figure}[h]
  \includegraphics[width=7cm]{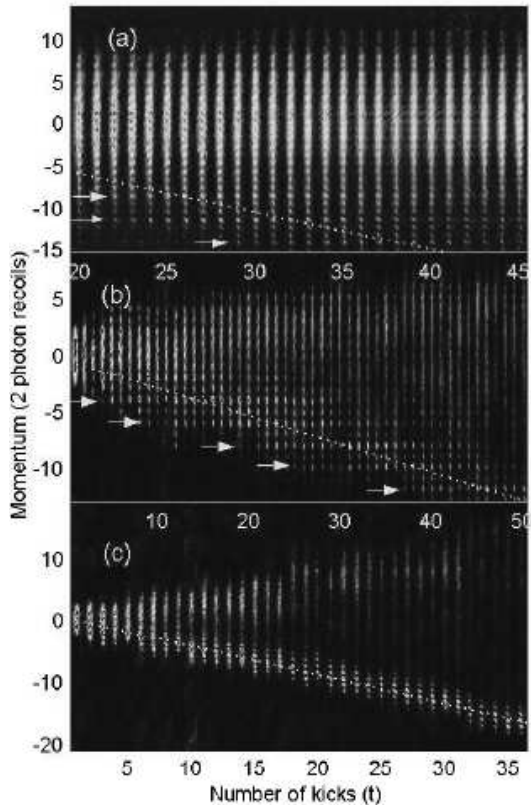}\\
  \caption{Horizontally stacked momentum distributions as a function of number of kicks ($t$)
  for (a) $T=22.68\ {\rm \mu s}$, which is close to (2/3)$T_{1/2}$,
  (b) $T=17.1\ {\rm \mu s}$, which is close to (1/2)$T_{1/2}$, and (c) $T=72.4\ {\rm \mu s}$
  which is close to 2$T_{1/2}$. Note the different axes
  for (a), (b) and (c). The arrows in (a) and (b) show that primarily orders
  separated by $b\hbar G$ participate in each of the QAMs.
  Dotted lines show the fit to $\epsilon$-classical theory of Eq.~(\ref{momentum}).}\label{kicks}
\end{figure}
To investigate the properties of higher order resonances of the
quantum $\delta$-kicked accelerator further, we conducted a series
of experiments where kick number was increased at a fixed kicking
period close to each of the above resonances. Figure~\ref{kicks}(a)
is a scan of number of kicks close to (2/3)$T_{1/2}$. This hints
that the QAM primarily consists of momentum states separated by $3
\hbar G$. The scan of kick number close to $(1/2)T_{1/2}$ of
Fig.~\ref{kicks}(b) shows much more clearly that the QAM is composed
of momentum states separated by 2$\hbar G$. These momentum states
are emphasized by arrows in Figs.~\ref{kicks}(a) and 2(b). In
contrast, at the Talbot time (2$T_{1/2}$), the QAM can include
neighboring momentum states as seen from Fig.~\ref{kicks}(c). This
behavior suggests that the QAM around the higher order resonances
can form whenever the momentum orders separated by $b \hbar G$
rephase during the time between the kicks. This is analogous to what
has been postulated to occur (but never directly observed) for the
kicked rotor resonances \cite{bil1}, and is consistent with what is
known of the fractional Talbot effect \cite{optics talbot}.

Given this structure for the QAMs it is now possible to explain why
an increased $\phi_d$ is needed to observe QAMs near resonances with
larger $b$. This is due to the fact that a kick diffracts atoms into
a wider range of momentum orders for larger $\phi_d$ so that the
orders separated by $b \hbar G$ can be more readily populated.
Recall that the population of a momentum state $l\hbar G$  is
proportional to $|J_l(\phi_d)|^2$. This in turn explains why it is
progressively more difficult to observe the QAMs and associated
resonances as $b$ becomes greater. Since $\phi_d$ must be increased
in this situation, the distribution of momentum states that do not
participate in a QAM broadens. This can mask the presence of a QAM
in either a scan of kick period or kick number especially in the
case of experiments with only a few kicks.

\begin{figure}[h]
  \includegraphics[width=7cm]{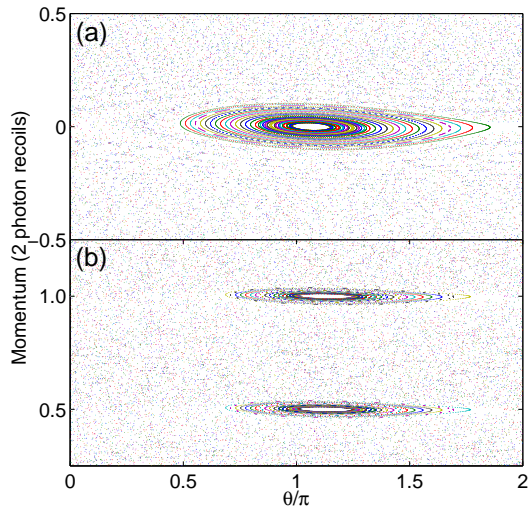}\\
  \caption{Phase space plot of the map from
  (\ref{map})
  for (a) $\phi_d=1.5$ and $T=17.1 \mu s$, which is close to (1/2) $T_{1/2}$, and
  (b) $\phi_d=1.8$ and $T=22.6 \mu s$ which is close to (2/3) $T_{1/2}$. The stable islands
  represent QAMs.}\label{map13by4}
\end{figure}

Although the $\epsilon$-classical model can not explain the momentum
state structure of a QAM as seen in Figs.~\ref{1by4period} and
\ref{kicks}, it predicts accurately the average momentum imparted to
a QAM. FGR attribute the existence of a QAM to the presence of
stable islands in the phase space map generated by Eq.\thinspace
(\ref{map}), here shown in Fig.\thinspace \ref{map13by4}. The
momentum axes in these maps are over a range of $1\hbar G$. The
appearance of two islands in Fig.~\ref{map13by4}(b) shows that the
repetition in initial momentum ($p_i$) for a QAM to appear at
(2/3)$T_{1/2}$ is $0.5\ \hbar G$. For the case of the kicked rotor,
at a resonance with a given $a$ and $b$ a quadratic growth of mean
energy occurs when the initial momentum is a multiple of $\hbar G/a$
\cite{qam_th}.  To calculate a similar result for the
$\delta$-kicked accelerator we can apply a method analogous to that
used to first understand the QAM at the primary quantum resonances
\cite{qam_th_gil}. This is done by setting the difference in phase
acquired by momentum states separated by $b \hbar G$ during the free
evolution between any two kicks to an integer multiple of $2\pi$.
This gives the condition on initial momentum (in unts of $\hbar G$)
at which a QAM appears as
\begin{equation}
p_i={2 \pi j \over \tau b}+{b \over 2}-{\eta \over 2},
\label{repeat}
\end{equation}
where $j$ is an integer. The last term here is small enough to be
ignored for the parameters used in our experiment. It can be seen
that the QAM in the $\delta$-kicked accelerator are spaced by
$\Delta p_i = 1/\tau b\approx1/a$.
\begin{figure}
  \includegraphics[width=7cm]{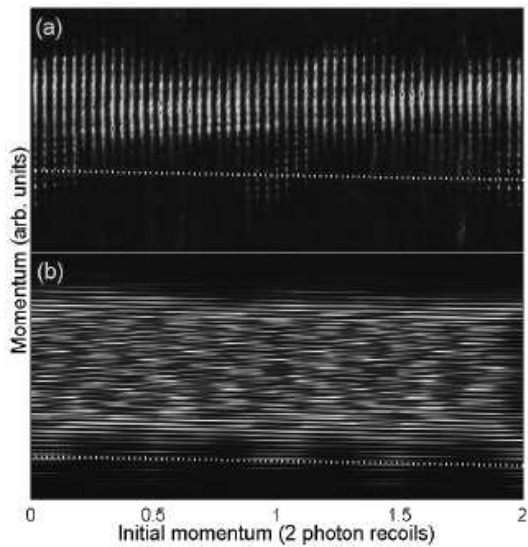}\\
  \caption{Horizontally stacked momentum distributions as a function on the initial momentum
  of the BEC before kicking with
  (a) 30 kicks with a period of 17.1 $\mu$s (b) a numerical simulation
  for 70 kicks with the period of 22.6 $\mu$s. The position of the modes are
  indicated by the dotted lines}\label{betascan}
\end{figure}
To observe the detailed phase space structure experimentally, it is
necessary that the momentum width should be much narrower than
$\hbar G$. In our experiments the BEC had a momentum width of 0.1
$\hbar G$ which makes it an excellent candidate for this task.
Figure \ref{betascan}(a) shows the results from experiments in which
the effective initial momentum of the BEC was changed by moving the
standing wave using a difference in frequency between the kicking
AOMs. The kicking period was near (1/2)$T_{1/2}$ in
Figs.~\ref{betascan}(a). The numerical simulation for (2/3)$T_{1/2}$
is displayed in Fig.~\ref{betascan}(b). Experiments in which the
initial momentum was scanned for (2/3)$T_{1/2}$ showed no distinct
features. This was most likely because the modes at this resonance
are very weak, as can be seen from Figs.~\ref{1by4period}(b) and
\ref{kicks}(a). The numerical simulations show observable modes
emerging after 70 kicks. Our experiments can not reach 70 kicks with
the large value of $\phi_d$ necessary to observe the mode at
(2/3)$T_{1/2}$. Nevertheless, the simulations do show that for
(2/3)$T_{1/2}$ the mode appears four times in a range of $2\hbar G$.
Thus, the initial momenta at which the modes can exist are separated
by $\hbar G$ in the case of $(1/2) T_{1/2}$ ($a=1$) and $\hbar G/2$
for $(2/3)T_{1/2}$ ($a=2$) in agreement with Eq.~(\ref{repeat}). The
periodicity of QAMs was deduced from the separation of QAMs in
Fig.~\ref{betascan}.

In conclusion, we have experimentally demonstrated the existence of
higher order resonances in the quantum $\delta$-kicked accelerator.
This was possible through the observation of QAMs near these
resonances. The $\epsilon$-classical theory of FGR was generalized
to predict the behavior of the system near the higher order
resonances. Furthermore, we were able to explore the phase space
structures produced by maps of the generalized theory. The momentum
transferred to a QAM at a resonance was in agreement with the
theory. The narrow momentum distribution of the BEC allowed us to
observe the momentum state structure of the QAMs. It was found that
QAMs near higher order resonances can have a very different
structure which is reminiscent of the momentum state structure
produced by the fractional Talbot effect. This work opens the door
towards the study of higher order QAMs near higher order resonances.
Other interesting questions include the effect of stronger mean
field interactions and the enhancement of QAMs using more complex
initial states.

\end{document}